\documentclass[sigconf,nonacm]{acmart}
\AtBeginDocument{%
  }

\setcopyright{acmlicensed}
\copyrightyear{2018}
\acmYear{2018}
\acmDOI{XXXXXXX.XXXXXXX}
\acmConference[Conference acronym 'XX]{Make sure to enter the correct
  conference title from your rights confirmation email}{June 03--05,
  2018}{Woodstock, NY}
\acmISBN{978-1-4503-XXXX-X/2018/06}

\newcommand{\vs}{{\it vs.} }

\newcommand{\stitle}[1]{\vspace{0.4ex}\noindent{\bf #1}}

\usepackage{multirow}
\usepackage{booktabs}
\usepackage{graphicx}
\usepackage{subcaption}

\begin{document}

\title{RGL: A Graph-Centric, Modular Framework for Efficient Retrieval-Augmented Generation on Graphs}

\author{Yuan Li}
\affiliation{%
  \institution{National University of Singapore}
  \country{Singapore}
}
\email{li.yuan@u.nus.edu}

\author{Jun Hu}
\affiliation{%
  \institution{National University of Singapore}
  \country{Singapore}
}
\email{jun.hu@nus.edu.sg}

\author{Jiaxin Jiang}
\affiliation{%
  \institution{National University of Singapore}
  \country{Singapore}
}
\email{jxjiang@nus.edu.sg}

\author{Zemin Liu}
\affiliation{%
  \institution{Zhejiang University}
  \country{China}
}
\email{liu.zemin@zju.edu.cn}

\author{Bryan Hooi}
\affiliation{%
  \institution{National University of Singapore}
  \country{Singapore}
}
\email{bhooi@comp.nus.edu.sg}

\author{Bingsheng He}
\affiliation{%
  \institution{National University of Singapore}
  \country{Singapore}
}
\email{hebs@comp.nus.edu.sg}

\renewcommand{\shortauthors}{Trovato et al.}

\begin{abstract}
Recent advances in graph learning have paved the way for innovative retrieval-augmented generation (RAG) systems that leverage the inherent relational structures in graph data. However, many existing approaches suffer from rigid, fixed settings and significant engineering overhead, limiting their adaptability and scalability. Additionally, the RAG community has largely overlooked the decades of research in the graph database community regarding the efficient retrieval of interesting substructures on large-scale graphs. In this work, we introduce the \textbf{R}AG-on-\textbf{G}raphs \textbf{L}ibrary (RGL), a modular framework that seamlessly integrates the complete RAG pipeline—from efficient graph indexing and dynamic node retrieval to subgraph construction, tokenization, and final generation—into a unified system. RGL addresses key challenges by supporting a variety of graph formats and integrating optimized implementations for essential components, achieving speedups of up to 143$\times$ compared to conventional methods. Moreover, its flexible utilities, such as dynamic node filtering, allow for rapid extraction of pertinent subgraphs while reducing token consumption. Our extensive evaluations demonstrate that RGL not only accelerates the prototyping process but also enhances the performance and applicability of graph-based RAG systems across a range of tasks.
\footnote{\url{https://github.com/PyRGL/rgl}}
\end{abstract}


\maketitle

\section{Introduction}

Recent advances in graph learning have witnessed an explosion of methods aimed at enhancing various facets of retrieval-augmented generation (RAG) on graphs~\cite{guo2024lightrag,hu2024grag,li2024subgraphrag}. Given a query, RAG retrieves relevant samples (context) from existing data and generates responses based on the retrieved information. Retrieval-augmented generation on graphs (RoG) extends RAG by leveraging graph structures to retrieve contextual information more effectively. Various applications on graphs, such as question answering, node classification, and recommendation---which contain rich structural data (e.g., user-item interactions~\cite{hu2024modalityindependentgraphneuralnetworks}, paper citation networks~\cite{DBLP:journals/tkde/HuHH24}, and more~\cite{zhang2025aggregate,zhang2024collaborate})---can potentially benefit from RoG techniques~\cite{he2025g,edge2024local}.

\begin{figure}[!tp]
    \centering
    \includegraphics[width=\linewidth]{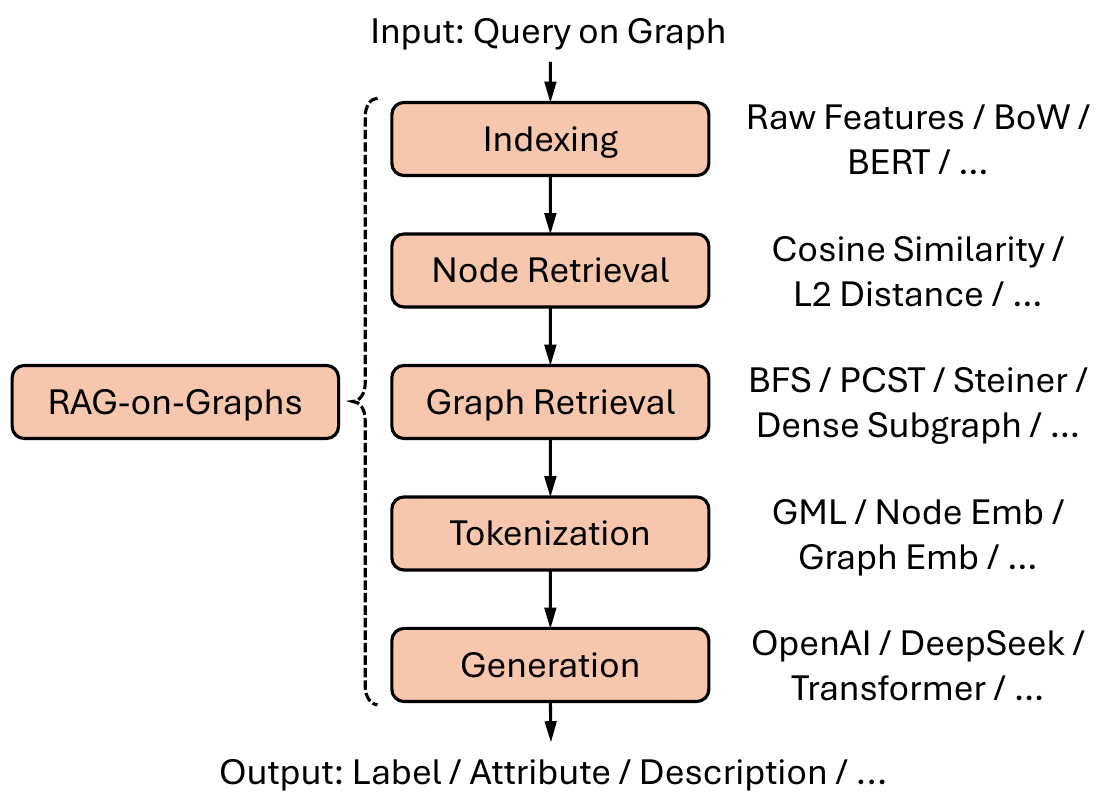}
    \caption{The pipeline of RAG-on-Graphs.}
    \label{fig:pipeline}
\end{figure}

\stitle{General RAG-on-Graph Pipeline.} Given a graph---such as a social network or an E-commerce graph---we illustrate a typical RAG-on-Graph pipeline in Figure~\ref{fig:pipeline}. The process begins with 1) \emph{Indexing}, where nodes are organized for efficient access. Next, 2) \emph{Node Retrieval} selects relevant nodes based on connectivity or attributes, after which 3) \emph{Graph Retrieval} constructs subgraphs to capture local structures. These subgraphs are then converted into a sequential format during 4) \emph{Tokenization}, rendering them compatible with state-of-the-art language models for the final 5) \emph{Generation} stage. This pipeline underpins more advanced integration of graph data into RAG workflows.

\begin{figure}[!tp]
    \centering
    \begin{subfigure}[b]{0.49\columnwidth}
        \centering
        \includegraphics[width=\columnwidth]{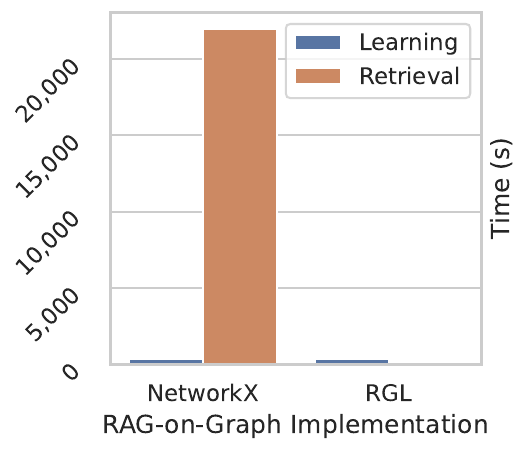}
        \caption{OGBN-Arxiv}
        \label{fig:Arxiv}
    \end{subfigure}
    \hfill
    \begin{subfigure}[b]{0.49\columnwidth}
        \centering
        \includegraphics[width=\columnwidth]{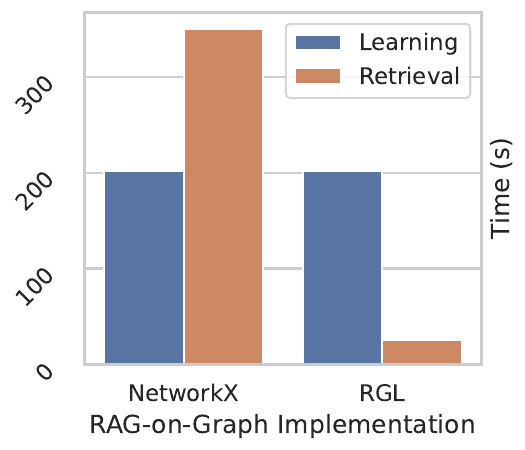}
        \caption{Sports}
        \label{fig:Sports}
    \end{subfigure}
\caption{Time consumption of different graph retrieval implementations. The learning time involves forward and backward propagation operations, while the retrieval time is introduced by RAG-on-Graph operations.}
\label{fig:time-q10000}
\end{figure}

Although the potential of RAG-on-Graphs is significant, its practical implementation remains challenging. First, many recent models are developed under fixed settings, limiting their adaptability to new datasets or the integration of novel components. For instance, GraphRAG~\cite{edge2024local} and LightRAG~\cite{guo2024lightrag} assume textual input for constructing knowledge graphs, which restricts their support for customized graphs---such as social networks or E-commerce graphs---and consequently limits their flexibility. Second, the requirement to implement each stage from scratch not only increases the implementation burden but also diverts researchers from focusing on methodological innovations. Finally, naive implementations of these stages can lead to efficiency pitfalls, particularly during the graph retrieval phase. This stage typically becomes a bottleneck, especially for large graphs, as illustrated by the NetworkX~\cite{hagberg2008exploring} implementations in Figure~\ref{fig:time-q10000}. More details are provided in the experimental section.

Given these challenges, efficient graph retrieval emerges as a critical component. Researchers have developed sophisticated algorithms~\cite{jiang2019generic,jiang2020ppkws,jiang2023dkws} and indexing methods to accelerate graph queries in complex domains such as social networks, bioinformatics, and knowledge graphs. However, the opportunity to integrate these advances with RoG has largely been overlooked by the broader RAG community.

To address these challenges and capitalize on the emerging opportunities, we introduce the \textbf{RAG-on-Graphs Library (RGL)}, a modular framework that integrates the complete pipeline. RGL overcomes the limitations of fixed settings and the heavy burden of building each component from scratch by providing a comprehensive data manager that supports various graph formats and by incorporating optimized graph indexing and retrieval algorithms (with key components implemented in C++). This design not only facilitates rapid prototyping but also effectively addresses the bottlenecks in graph retrieval, as evidenced by up to 143$\times$ speedups over existing implementations (see Figure~\ref{fig:time-q10000}). Additionally, RGL offers flexible utilities, such as dynamic node filtering, which enable researchers to quickly extract the most relevant subgraphs and reduce token consumption during generation, thereby directly addressing the issues discussed above.

The remainder of this paper is organized as follows. 
Section~\ref{sec:overview} details the methodology employed in this research, explicating the library design, data collection, and analysis techniques. 
In Section~\ref{sec:experiments}, the results of the study are presented, accompanied by relevant tables and figures to facilitate understanding. 
Finally, Section~\ref{sec:conclusion} concludes the paper by summarizing the key outcomes, acknowledging the limitations, and proposing directions for future research.

\section{RGL Overview}
\label{sec:overview}

\begin{figure}[!tp]
    \centering
    \includegraphics[width=\linewidth]{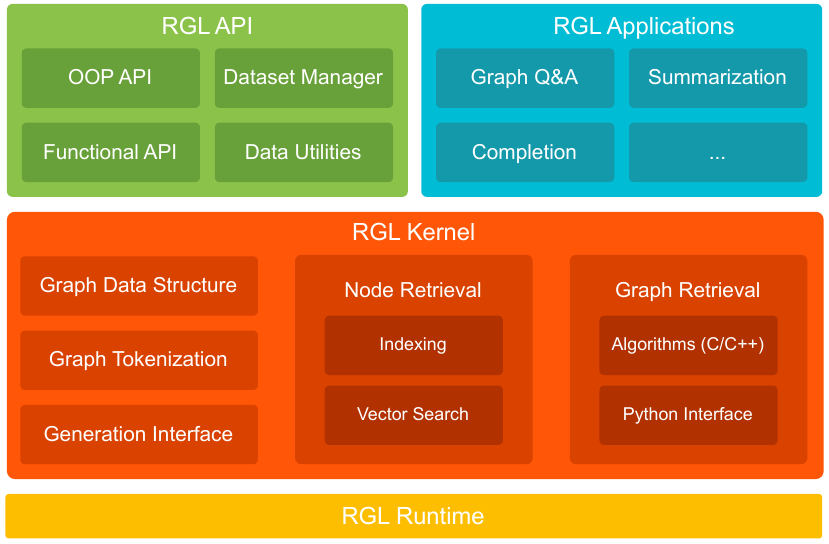}
    \caption{An overview of the RGL toolkit.}
    \label{fig:components}
\end{figure}

RGL is a modular toolkit designed to streamline the development of RAG techniques on graph data. As illustrated in Figure~\ref{fig:components}, RGL is composed of four primary components---Runtime, Kernel, API, and Applications---each providing specialized functionalities for efficient and flexible RAG-on-Graphs workflows.

\subsection{RGL Kernel}
\label{subsec:rgl-kernel}

The RGL Kernel provides fundamental components that handle graph data, retrieval, and generation processes. These components are carefully optimized to support various RAG scenarios, including indexing, high-performance retrieval, and batch processing. 

\subsubsection{Graph Data Structure.}
RGL provides an intuitive Python interface for constructing and manipulating graph structures. Researchers can effortlessly build RGL graph objects using native Python objects. In addition, RGL ensures seamless conversions to and from popular frameworks such as DGL~\cite{wang2019deep} and PyTorch Geometric (PyG)~\cite{fey2019fast}, allowing users to leverage the rich datasets available in these libraries.

\subsubsection{Node Retrieval.}
To facilitate semantic-level graph querying, RGL provides indexing and vector search utilities. Graph nodes and edges can be embedded into semantic vectors, enabling similarity-based retrieval that goes beyond simple keyword or ID lookups.

\subsubsection{Graph Retrieval.}
RGL implements a suite of \textbf{efficient graph retrieval algorithms} that leverage Python's ease-of-use features---empowered by extensive libraries like PyTorch and DGL---and C++ efficient implementations, all connected via pybind11 bindings. This hybrid approach enables computationally intensive tasks---such as shortest-path computations, neighbor expansions, and subgraph extractions---to be offloaded to optimized C++ routines, resulting in performance improvements that significantly surpass those of Python-based libraries like NetworkX~\cite{hagberg2008exploring}. By batching operations, RGL reduces function call overhead and increases throughput, making it well-suited for large-scale graph processing tasks.

\subsubsection{Generation Interface.}
This interface bridges the gap between retrieved subgraphs and downstream language models. It handles tokenization, prompt construction, and generation calls.

\subsection{RGL Runtime}
\label{subsec:rgl-runtime}

The RGL Runtime manages resource allocation, caching, and parallelization across kernel components, abstracting distributed execution and memory management to ensure scalability and performance. It also integrates with popular graph learning frameworks such as DGL~\cite{wang2019deep} and PyTorch Geometric (PyG)~\cite{fey2019fast}, enabling seamless incorporation of specialized GNN layers and operations.

\subsection{RGL API}
\label{subsec:rgl-api}

RGL offers a dual-API approach---an \textbf{Object Oriented Programming (OOP) API} and a \textbf{Functional API}---to accommodate a wide range of development styles and use cases:

\subsubsection{OOP API}
The OOP API provides class-level interfaces for constructing, training, and deploying RAG workflows. These classes encapsulate data structures, retrieval logic, and generation calls.

\subsubsection{Functional API}
For more fine-grained control, the Functional API exposes key operations (e.g., subgraph extraction, embedding, tokenization) as composable functions. This design is especially useful for advanced scenarios, such as meta-learning or dynamic parameterization, where developers may need to inject custom logic at various stages of the pipeline.

\subsubsection{Dataset Manager \& Utilities.}
RGL includes utilities for handling various graph formats, loading and preprocessing data, and managing node or edge attributes. This streamlines the process of experimenting with new datasets and graph structures, reducing boilerplate code and speeding up development.

\subsection{RGL Applications}
\label{subsec:rgl-applications}

Built on top of the kernel, runtime, and APIs, \textbf{RGL Applications} serve as end-to-end solutions for common tasks in graph-based RAG. Along with the open-source library, we provide the following demonstrative examples:

\begin{itemize}
    \item \stitle{Completion:} Enhance data completion tasks using RGL by retrieving graph contexts to infer missing data more effectively. The framework’s advanced analytics capabilities provide comprehensive insights into graph structure, which aids in accurate prediction and completion of incomplete datasets, thus bolstering model accuracy.

    \item \stitle{Summarization:} Utilize RGL for graph-based content summarization by employing subgraph extraction and generation models. RGL's efficient graph algorithms allow for fast identification of pivotal subgraph components, enabling thorough summarization strategies to organically generate concise summaries.

    \item \stitle{Graph Q\&A:} Implement node- and graph-level question answering using RAG-on-Graphs by integrating the RGL framework. The RGL framework facilitates the extraction of relevant graph information for real-time question answering, supporting both intrinsic graph queries and comprehensive node-centric inquiries.
\end{itemize}

Developers can customize these applications or create entirely new ones by combining the RGL kernel modules with API interfaces. The result is a powerful, extensible toolkit that simplifies the entire lifecycle of RAG-on-Graphs applications.

\section{Empirical Evaluation}
\label{sec:experiments}

In this section, we present a empirical evaluation across two challenging tasks: modality completion and abstract generation. By simulating realistic scenarios---ranging from sparse modality data in recommendation settings to prompt-driven text generation---we empirically show the efficacy of RAG-on-Graphs on various graph learning tasks. 

\subsection{Efficiency}

\begin{figure}[!tp]
    \centering
    \begin{subfigure}[t]{0.49\columnwidth}
        \centering
        \includegraphics[width=\linewidth]{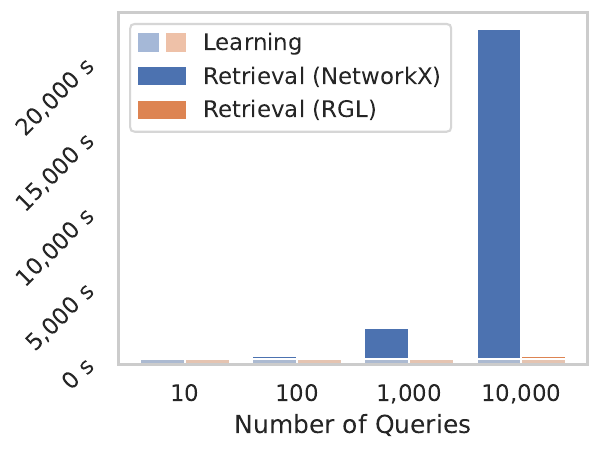}
        \caption{BFS (OGBN-Arxiv)}
    \end{subfigure}
    \hfill
    \begin{subfigure}[t]{0.49\columnwidth}
        \centering
        \includegraphics[width=\linewidth]{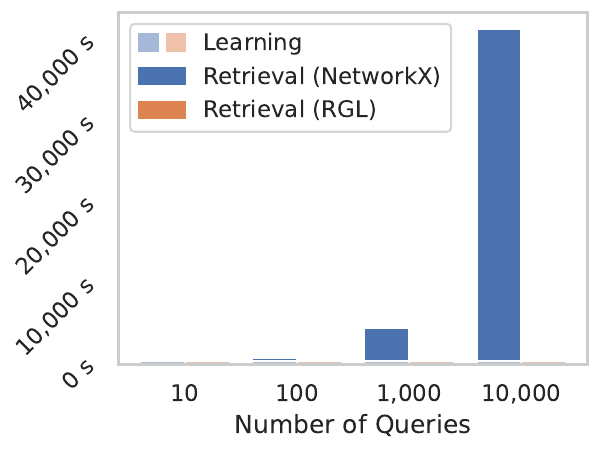}
        \caption{Steiner (OGBN-Arxiv)}
    \end{subfigure}
    
    \vspace{0.5cm}
    
    \begin{subfigure}[t]{0.49\columnwidth}
        \centering
        \includegraphics[width=\linewidth]{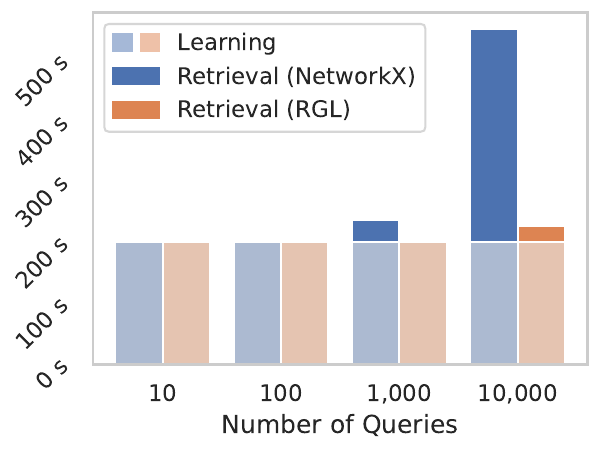}
        \caption{BFS (Sports)}
    \end{subfigure}
    \hfill
    \begin{subfigure}[t]{0.49\columnwidth}
        \centering
        \includegraphics[width=\linewidth]{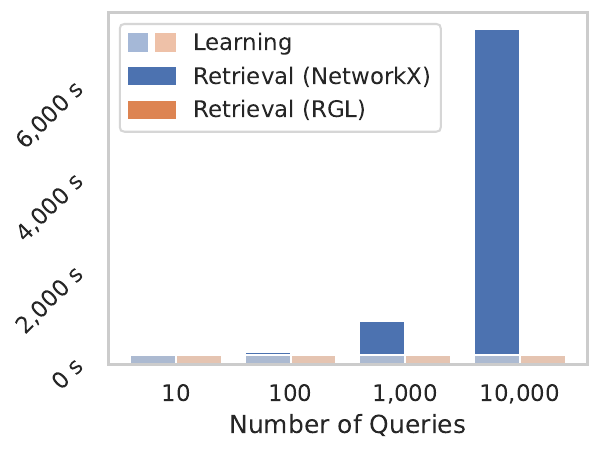}
        \caption{Steiner (Sports)}
    \end{subfigure}

    \caption{Time consumptions (s) \vs query counts across graph retrieval methods and datasets. The light colors denote the original training time, while the dark colors mark the additional time associated with graph retrieval.}
    \label{fig:time}
\end{figure}

Figure~\ref{fig:time} reports the time consumption under different numbers of queries, where in our settings a \textbf{query} involves the retrieval process for a certain node. We compare the standard graph library NetworkX~\cite{hagberg2008exploring} with algorithms implemented in RGL. The time consumption is separated into two components: 1) The learning time, which is consistent for a given dataset and typically involves the forward and backward computations; and 2) The retrieval process, which is an additional stage that augments the learning process with the retrieved contexts for the query nodes.

\textbf{NetworkX suffers from steep retrieval costs.} NetworkX's retrieval time grows dramatically with the number of queries. For example, on OGBN-Arxiv the baseline Steiner graph takes more than \textbf{11 hours} to process 10,000 queries, rendering it infeasible for large-scale scenarios.

\textbf{RGL offers efficient large-scale retrieval.} RGL consistently exhibits short retrieval times, incurring only a minor additional overhead compared with the learning time. Specifically, RGL completes the same 10,000 queries on OGBN-Arxiv in under \textbf{5 minutes}, indicating a drastic improvement compared with baselines.

These experiments confirm that NetworkX becomes prohibitively expensive beyond a few hundred queries, whereas RGL scales more gracefully. Given these observations, we adopt RGL for all subsequent performance evaluations, as its total runtime remains manageable for large-scale graph retrieval tasks.

\subsection{Performance}

\subsubsection{Modality Completion}
In this section, we evaluate the performance of multi-modality completion on two challenging multimodal recommendation datasets. The goal is to recover missing modality-specific features, which is essential for enhancing downstream recommendation tasks when data is sparse or incomplete.

\stitle{Dataset Statistics.} 
We evaluate our approach on two bipartite graphs with multimodal data. The \textit{Baby} dataset comprises 19,445 users and 7,050 items, resulting in 160,792 recorded interactions. In contrast, the \textit{Sports} dataset includes 35,598 users and 18,357 items with 296,337 interactions. 

\stitle{Baselines.} 
Our experimental setup compares several completion methods. Baseline techniques include Fill0, NeighMean \cite{malitesta2024we}, PPR \cite{malitesta2024we}, Diffusion \cite{li2025generating}, kNN, and kNN-Neigh. In addition, we propose three variants of the RGL method based on different subgraph construction strategies: RGL-Steiner, RGL-Dense, and RGL-BFS. 

\stitle{Evaluation.} 
We use the public data splits including training, validation, and test sets, following prior works~\cite{hu2024modality,zhou2023tale}. We simulate the missing modality scenarios by randomly masking a subset of the features during training. We follow prior work~\cite{li2025generating} to set the missing rate to 40\%, underscoring the importance of effective modality completion with sparse modality data. The recommendation performance is measured using Recall at 20 (R@20) and Normalized Discounted Cumulative Gain (N@20). We repeat all experiments 5 times on a V100-32GB GPU and report the mean scores.

\stitle{Results and Analysis.}
Table~\ref{tab:completion} summarizes the performance of our method under varying missing rates and completion strategies. Our approach consistently outperforms all baselines across the datasets. In particular, the RGL-based subgraph construction methods (BFS, Dense, Steiner) yield the best performance in both recall and NDCG scores on all datasets. These findings validate the effectiveness of leveraging RAG-on-Graph techniques for multi-modality completion in sparse data scenarios.

\begin{table}[!tp]
\centering
\caption{Modality completion performance with different missing rates (MR) and completion methods (Compl.). RGL-BFS/Dense/Steiner denote different subgraph construction methods using the retrieved nodes.}
\label{tab:completion}
\begin{tabular}{@{}c|cccc@{}}
\toprule
\multirow{2}{*}{Method} & \multicolumn{2}{c}{Baby} & \multicolumn{2}{c}{Sports} \\ \cmidrule(l){2-5} 
 & R@20 & \multicolumn{1}{c|}{N@20} & R@20 & N@20 \\ \midrule
Fill0 & 0.0902 & \multicolumn{1}{c|}{0.0393} & 0.0972 & 0.0434 \\
NeighMean \cite{malitesta2024we} & 0.0890 & \multicolumn{1}{c|}{0.0393} & 0.0997 & 0.0445 \\
PPR \cite{malitesta2024we} & 0.0906 & \multicolumn{1}{c|}{0.0395} & 0.0977 & 0.0439 \\
Diffusion \cite{li2025generating} & 0.0746 & \multicolumn{1}{c|}{0.0325} & 0.0860 & 0.0384 \\
kNN & 0.0924 & \multicolumn{1}{c|}{\textbf{0.0405}} & 0.0993 & 0.0446 \\
kNN-Neigh & 0.0902 & \multicolumn{1}{c|}{0.0393} & 0.0987 & 0.0444 \\ \midrule
RGL-Steiner & \textbf{0.0936} & \multicolumn{1}{c|}{\textbf{0.0405}} & 0.1004 & 0.0449 \\
RGL-Dense & 0.0932 & \multicolumn{1}{c|}{\textbf{0.0405}} & \textbf{0.1005} & 0.0448 \\
RGL-BFS & 0.0928 & \multicolumn{1}{c|}{\textbf{0.0405}} & 0.1003 & \textbf{0.0450} \\ \bottomrule
\end{tabular}
\end{table}

\subsubsection{Abstract Generation} 
In this section, we compare abstract generation approaches across context construction methods and language models, demonstrating the effectiveness of RGL.

\begin{table}[!tp]
\centering
\caption{Abstract generation performance across different models and prompted contexts.}
\label{tab:absgen}
\begin{tabular}{@{}cccc@{}}
\toprule
\multicolumn{1}{c|}{\multirow{2}{*}{Method}} & \multicolumn{3}{c}{OGBN-Arxiv to Arxiv2025} \\ \cmidrule(l){2-4} 
\multicolumn{1}{c|}{} & ROUGE-1 & ROUGE-2 & ROUGE-L \\ \midrule
\multicolumn{4}{c}{GPT-4o-mini} \\ \midrule
\multicolumn{1}{c|}{SelfNode} & 0.3791 & 0.0754 & 0.1775 \\
\multicolumn{1}{c|}{kNN} & 0.3814 & 0.0758 & 0.1796 \\
\multicolumn{1}{c|}{RGL-Steiner} & \textbf{0.3831} & \textbf{0.0771} & 0.1796 \\
\multicolumn{1}{c|}{RGL-Dense} & 0.3789 & 0.0720 & 0.1790 \\
\multicolumn{1}{c|}{RGL-BFS} & 0.3815 & 0.0763 & \textbf{0.1801} \\ \midrule
\multicolumn{4}{c}{DeepSeek-V3} \\ \midrule
\multicolumn{1}{c|}{SelfNode} & 0.3754 & 0.0782 & 0.1806 \\
\multicolumn{1}{c|}{kNN} & 0.3717 & 0.0762 & 0.1828 \\
\multicolumn{1}{c|}{RGL-Steiner} & 0.3786 & 0.0790 & 0.1825 \\
\multicolumn{1}{c|}{RGL-Dense} & \textbf{0.3817} & 0.0784 & 0.1855 \\
\multicolumn{1}{c|}{RGL-BFS} & 0.3804 & \textbf{0.0802} & \textbf{0.1859} \\ \bottomrule
\end{tabular}
\end{table}

\stitle{Dataset Statistics.} 
For abstract generation, we leverage a large-scale citation network extracted from OGBN-Arxiv, which comprises 169,343 nodes, 1,157,799 edges, 128-dimensional features, and 40 classes. These real-world data pose a demanding task of synthesizing concise, informative abstracts from complex graph structures and textual information.

\stitle{Baselines and Prompted Contexts.} 
In addition to our proposed RGL variants (RGL-Steiner, RGL-Dense, and RGL-BFS), we consider two baselines: SelfNode and kNN. Furthermore, we evaluate the generation quality with two different large language models (LLMs): GPT-4o-mini and DeepSeek-V3.

\stitle{Evaluation.} 
We inspect a zero-shot transfer scenario---OGBN-Arxiv to Arxiv2025---that occurs after the LLM knowledge cutoff dates (October 1, 2023 for GPT-4o-mini, and July 1, 2024 for DeepSeek-V3) to avoid knowledge leakage. We employ ROUGE-1, ROUGE-2, and ROUGE-L~\cite{lin2004rouge} as our primary evaluation metrics, which respectively quantify the overlap of unigrams, bigrams, and longest common subsequences between generated and reference abstracts. These metrics provide insights into content fidelity at different levels of granularity.

\stitle{Results and Analysis.} 
Table~\ref{tab:absgen} summarizes the performance of our methods on the OGBN-Arxiv to Arxiv2025 task. The key findings are as follows:
\begin{itemize}
    
\item When utilizing the GPT-4o-mini model, RGL-Steiner achieves the highest ROUGE-1 and ROUGE-2 scores, whereas RGL-BFS leads in ROUGE-L. In contrast, with the DeepSeek-V3 model, RGL-Dense attains the top ROUGE-1 score, and RGL-BFS continues to excel in ROUGE-L.

\item These results demonstrate that our RGL framework effectively leverages both graph structure and contextual cues, thereby producing abstracts that are both coherent and highly representative of the source content.
\item The variance in performance across different graph traversal strategies (Steiner, BFS, Dense) with varied modeling techniques (GPT-4o-mini, DeepSeek-V3) suggests that the RGL framework's adaptability is crucial for optimizing abstract generation tasks.
\end{itemize}

\section{Conclusions}
\label{sec:conclusion}

In this paper, we introduced the RAG-on-Graphs Library (RGL), a modular and highly adaptable toolkit designed to streamline the integration of graph data into retrieval-augmented generation systems. Our experimental results, spanning modality completion and abstract generation tasks, convincingly demonstrate that RoG enhances graph learning performance. Specifically, it delivers notable speedups in graph retrieval processes and markedly improves the quality of the generated content. By integrating optimized graph processing techniques, providing flexible APIs, and ensuring seamless interfacing with state-of-the-art language models, RGL establishes a solid platform for advancing research in RoG applications. 

Looking ahead, several avenues for future work can be identified. Expanding the library to include a broader range of examples can facilitate better understanding and implementation. Furthermore, efforts to enhance user-friendliness will make RGL more accessible to a wider audience. Large-scale testing is necessary to further validate the robustness and scalability of the library. Additionally, exploring integration with other graph database tools could provide insightful synergies, thereby expanding RGL's applicability in complex graph environments.

\bibliographystyle{ACM-Reference-Format}
\bibliography{sample-base}

\end{document}